\documentclass[11pt]{article}
\usepackage{moriond,epsfig}

\def\gtsim{\raise 2pt \hbox {$>$} \kern-1.1em \lower 4pt \hbox {$\sim$}}

\def\be{\begin{equation}}
\def\ee{\end{equation}}
\def\bea{\begin{eqnarray}}
\def\eea{\end{eqnarray}}

\begin{document}
\vspace*{4cm}
\title{RADIO OBSERVATIONS OF GALAXY CLUSTERS: \\CONNECTION TO CLUSTER MERGERS}

\author{ L. FERETTI }

\address{Istituto di Radioastronomia INAF, Via P. Gobetti 101\\
40129 Bologna, Italy}

\maketitle\abstracts{
The main component of the intracluster medium (ICM) in clusters of
galaxies is represented by the X-ray emitting thermal plasma.  In
addition, the presence of relativistic electrons and large-scale
magnetic fields in a fraction of galaxy clusters
is demonstrated by the detection of large-scale synchrotron radio
sources, which have no optical counterpart and no obvious connection
to the cluster galaxies.
Observational results provide evidence that these
phenomena are related to cluster merger activity, which supplies the
energy for the reacceleration of the radio emitting particles. The
investigation of the halo-merger connection is of great importance to
the knowledge of the formation and evolution of clusters of galaxies.
}

\section{Introduction}

Clusters are formed by hierarchical structure formation processes.  In
this scenario, smaller units formed first and merged to larger and
larger units in the course of time.  The merger activity appears to be
continuing at the present time, and explains the relative abundance of
substructures found from X-ray and optical studies,
and the gas temperature gradients detected 
in rich clusters from  X-ray observations.  
The ICM in merging clusters is likely to be
in a violent or turbulent dynamical state.  It is found that a fraction of
clusters which have recently undergone
a merger event show diffuse radio sources, 
thus leading to the idea that these sources are energized by
turbulence and shocks in cluster mergers. 
In this paper, the link between diffuse 
radio sources and  cluster merger processes is discussed, with particular
emphasis on the observational aspect.

\section {Radio halos and relics}

The Coma cluster is the first cluster where a radio halo (Coma C) and
a relic (1253+275) have been detected (Willson 1970, Ballarati et
al. 1981).  
The radio halos are permeating the cluster
central regions, with a brightness
distribution similar to that of the X-ray gas (Govoni et al. 2001)
and a typical extent of \gtsim~1 Mpc. They are characterized by  steep
radio spectrum.  Limits of a few percent  have been
derived for their polarized emission.  
Relic sources are similar to halos in their low surface
brightness, large size and steep spectrum, but they are typically
found in cluster peripheral regions. 
Unlike halos, relics are highly polarized ($\sim$ 20 $-$ 30\%).

The number of clusters with halos and relics is presently around 50. 
Examples of radio images of cluster halos and relics are given in Fig. 1.
Their
properties have been recently reviewed by Giovannini \& Feretti (2002, 2004)
and Feretti (2005).  
From radio data, under equipartition conditions, the minimum energy content 
in halos and relics is of about 10$^{60}$ $-$ 10$^{61}$
erg, with  minimum non-thermal energy densities of 10$^{-14}$ $-$ 10$^{-13}$
erg cm$^{-3}$.  The last values are about 1000 times lower than 
values of the energy density of the thermal X-ray gas.  Thus, the
non-thermal components are not contributing to a significant fraction
of the mass or energy of a cluster.  However, the presence of relativistic
particles and large-scale magnetic fields are important for a
comprehensive physical description of the ICM in
galaxy clusters.

\begin{figure}
\centerline{
\psfig{figure=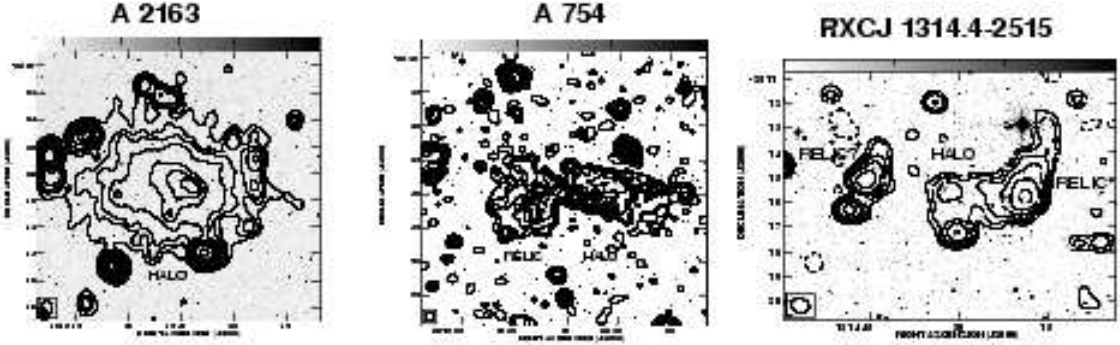,width=15.5cm}}
\caption
{Examples of diffuse radio sources in clusters, overlaid onto the optical
cluster image:
the giant radio halo in A~2163 (Feretti et al. 2001) is one of the most 
powerful known halos; the 
emission in A~754 (Bacchi et al. 2003) is quite irregular, including
a radio halo and a radio relic;  
the  complex emission in the cluster $RXCJ1314.4-2515$
(Feretti et al. 2005) is dominated by a central halo and a western
relic possibly
connected to each other, plus a possible second relic located on the
eastern side of the cluster.
}
\label{diff}
\end{figure}

\section {Connection to cluster merging processes}

All clusters hosting halos and relics are characterized by dynamical activity
related to merging processes.  
These clusters indeed show: i) substructures and
distortions in the X-ray brightness distribution (Schuecker et
al. 2001); ii) temperature gradients (Govoni et al. 2004) and gas
shocks (Markevitch et al. 2003a); iii) absence of a strong cooling
flow (Schuecker et al. 2001); iv) values of spectroscopic $\beta$ on
average larger than 1 (Feretti 2002); v) core radii significantly
larger than those of clusters classified as single/primary (Feretti
2002); vi) large distance from the nearest neighbours compared to
clusters with similar X-ray luminosity (Schuecker \& B\"ohringer
1999);  the latter fact supports the idea that
recent merger events lead to a depletion of the nearest neighbours.
Buote (2001) derived a correlation between the radio power of halos and
relics and the dipole power ratio of the cluster two-dimensional
gravitational potential.  Since power ratios are closely related
to the dynamical state of a cluster, this correlation represents
the first attempt to quantify the link between diffuse sources and
cluster mergers.

\begin{figure}
\centerline{
\psfig{figure=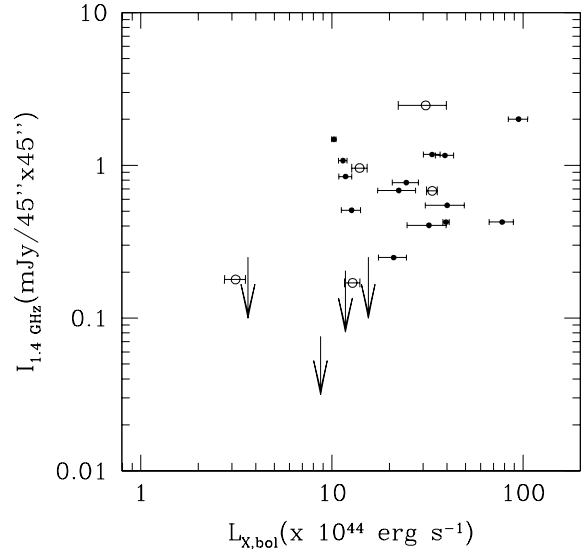,width=10cm}}
\caption
{Radio brightness of halos at 1.4 GHz versus the cluster bolometric
X-ray luminosity. Empty circles represent halos of size smaller than 1 Mpc. 
Arrows indicate upper limits obtained for merging clusters
where no diffuse radio halo has been detected.
}
\label{diff}
\end{figure}

The link between cluster merger processes and the presence of
radio halos and relics indicates a physical connection
between the thermal and non-thermal ICM components.
However, not all clusters showing merging processes are known to host
halos and/or relics.
This could be due to the limited sensitivity of current
radio telescopes. Indeed it is found that
the most powerful radio halos and relics are detected
in the most X-ray luminous clusters (see e.g. Feretti 2005). 
For halos, this trend is reflected in the
correlation of the radio brightness versus X-ray 
luminosity,  presented in Fig. 2. 
The  upper limits shown in the
plot are consistent with the correlation suggesting that clusters of low
X-ray luminosity might host faint diffuse sources. 
 On the other hand,
it is possible that giant halos are only present in the most X-ray
luminous clusters, i.e. above a treshold of X-ray luminosity (see
discussion in Bacchi et al. 2003). This would be expected in the framework
 of electron
reacceleration models (Cassano et al. 2006).  
Future radio data with next generation
instruments (LOFAR, LWA, SKA) will allow the detection of low
brightness/low power large halos, in order to clarify if halos are
present  in all merging clusters or only in the most massive ones.

\section {Spectral index distribution of radio halos}

A direct confirmation that the cluster merger supplies energy
to the radio halo can be obtained from maps of the radio
spectral index and their comparison to the
X-ray properties.
The radio spectral index is related to the energy and ageing of 
relativistic electrons, and
to the strength of the magnetic field in which they emit. These
parameters are both influenced by a cluster merger.
Thus, by combining high resolution spectral information and
X-ray images, it is possible to study the thermal-relativistic plasma
connection both on small scales (e.g. spectral index variations
vs. clumps in the ICM distribution) and on the large scale
(e.g. radial spectral index trends).

Feretti et al. (2004) obtained maps of the
radio spectral index between 0.3 and 1.4 GHz in A~665 and A~2163, 
and found  that regions characterized by 
flatter radio spectra appear to trace the
geometry of recent merger activity.  A similar behaviour is derived
in A~2744 (Orr\'u et al. 2006). Since regions of flatter spectra are
indication of the presence of more energetic radiating particles,
the above results support the idea that the merger process supplies energy
to the radio emitting electrons. In particular, one can compute 
that in regions of identical volume and same brightness at 0.3
GHz, a flattening of the spectral index from 1.3 to 0.8 implies 
an amount of energy of the electron population larger by a factor of
$\sim$ 2.5.
The difference in the spectral index can also be understood in terms of
electron ageing, i.e. by considering that a flatter spectrum 
generally reflects
 a spectral cutoff occurring at higher energies.  The implication
is that the electrons in the flat spectrum regions would have been
reaccelerated more recently.

The above results prove that the radio spectral index can be a powerful
tracer of the current physical properties of clusters, and confirm the
importance of cluster merger in the energetics of relativistic
particles responsible for the halo radio emission.

\section{Origin of radio halos}

The relativistic particles radiating in halos are 
likely to originate from AGN activity
(quasars, radio galaxies, etc.), or from star formation
(supernovae, galactic winds, etc.), or from the thermal pool
during violent processes connected to the cluster dynamical history.
Because of radiation losses, the radiating  particles have  short lifetimes,
of the order of 0.1 Gyr. This implies that they can travel a maximum
distance of $\sim$ 100 kpc during their lifetime. Given the
large size of the radio emitting regions, the relativistic particles
need to be reaccelerated by some mechanism, acting with an efficiency
comparable to the energy loss processes  (Petrosian 2001). 
It is found from observational results that a cluster merger 
plays a crucial role in the energetic of radio halos. 
Energy can be transferred from the ICM thermal component to the
non-thermal component through two possible basic mechanisms: 1)
acceleration at shock waves (Keshet et al. 2004); 2)
resonant or non-resonant interaction of electrons with MHD turbulence
(Brunetti et al. 2001, Fujita et al. 2003).

Shock acceleration is a first-order Fermi process of great importance
in radio astronomy, as it is recognized as the mechanism responsible
for particle acceleration in the supernova remnants.  The acceleration
occurs diffusively, in that particles scatter back and forth across
the shock, gaining at each crossing and recrossing an amount of energy
proportional to the energy itself.  The acceleration efficiency is
mostly determined by the shock Mach number.  
Based on observational evidences, it seems difficult to associate giant halos
to merger shocks, since the shocks are localized while the radio emission
of halos is generally much more extended.
In addition, Gabici \& Blasi (2003) argue that only shocks of high Mach number
(\gtsim~ 3) are suitable for the electron reacceleration,
whereas shocks detected so far with 
Chandra at the center of several
clusters have inferred Mach
numbers in the range of $\sim$ 1 - 2.5 (see e.g. Markevitch et al. 2003a).
Finally, the radio spectral
index distribution in A~665 (Feretti et al. 2004) shows no evidence of
radio spectral flattening at the location of the hot shock detected by
Chandra (Markevitch \& Vikhlinin 2001).

Therefore, although it cannot be excluded that shock acceleration may be
efficient in some particular regions of a halo (e.g. in A~520,
Markevitch et al. 2005), current observations globally
favour the scenario that  turbulence following a cluster shock,
rather than a shock itself, might be the major
mechanism responsible for the supply of energy to the electrons
radiating in radio halos.  Numerical simulations indicate that mergers
can generate strong fluid turbulence on scales of 0.1 - 1 Mpc.
The time during which the process is effective is of $\sim$ 10$^8$ 
years, so that the emission is expected to correlate with the
most recent or ongoing merger event. 
The emerging scenario is that turbulence reacceleration is the likely
mechanism to supply energy to the radio halos.  Recent theoretical developments
of this aspect can be found in Blasi 2004, Brunetti et al. (2004),
Cassano \& Brunetti (2005).

\begin{figure}
\centerline{
\psfig{figure=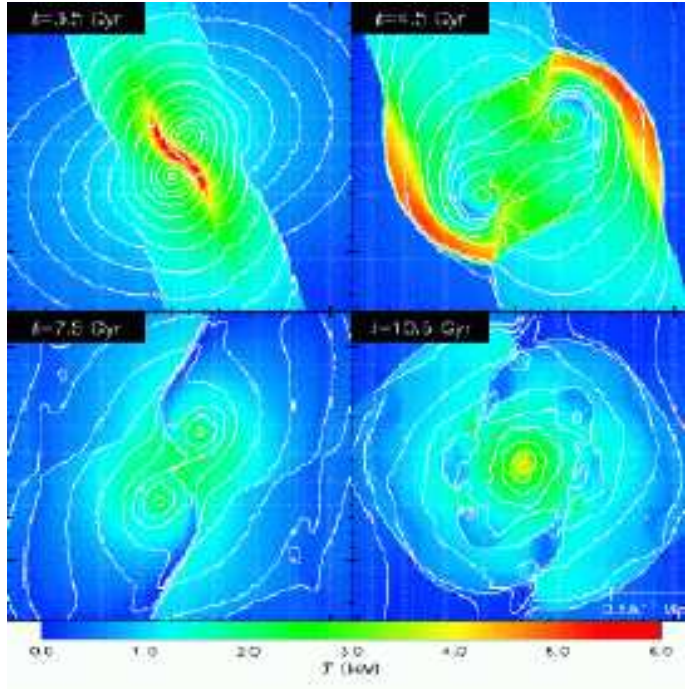,width=9cm}}
\caption
{Results of hydrodynamical simulations of a symmetric off-center merger by 
Ricker \& Sarazin (2001). The colors show the temperature, while the contours
represent the X-ray surface brightness. A shocked region is initially at
the cluster center, then shocks propagate to the outer part of the cluster.
}
\label{diff}
\end{figure}

\section {Origin of radio relics}

Current theoretical models predict that relativistic particles
radiating in radio relics are powered by energy dissipated in shock
waves produced  in the ICM during the
cluster formation history. This picture is supported
by numerical simulations on cluster mergers (Ricker \& Sarazin
2001, Ryu et al. 2003),
which predict that shocks forming at the cluster center at the early
stages of a cluster merger further 
propagate to the cluster periphery (Fig. 3).  

Two models have been proposed for the origin of the
relic radio emission. It could 
result from Fermi-I diffusive shock acceleration which produces
relativistic electrons from the ICM electrons, 
or it could be due to fossil radio bubbles, 
related to former active
radio galaxies, 
that are compressed by the passing shock wave and thus induced to emit
observable synchrotron emission again (Ensslin \& Gopal-Krishna 2001,
Ensslin \& Br\"uggen 2002, Hoeft et al. 2004).  
In both case, because of the
electron short radiative lifetimes, radio emission is produced close
to the location of the shock waves.  
This is consistent with
the relic elongated structure, almost perpendicular to the merger axis.

The detection of shocks in the cluster outskirts is presently very
difficult because of the very low X-ray brightness in these
regions. The X-ray data for radio relics are indeed very scarce. The
Chandra data of A~754 (Markevitch et al. 2003b) indicate that the
easternmost  boundary of the relic (see Fig. 1) coincides with
a region of hotter gas. From XMM  data of the same cluster,
Henry et al. (2004) show that the diffuse radio sources
(halo + relic) appear to be associated with high pressure regions.

A recent study of the  region of the radio relic 1253+275 
in the Coma cluster has been performed 
by Feretti \& Neumann (2006) using XMM-Newton data.
X-ray emission is detected at the location of the radio relic
(Fig. 4) and is found to be of thermal origin, 
connected to the sub-group around NGC~4839.  The best-fit gas
  temperature in the region of the relic and in its vicinity is in the
  range 2.8 $-$ 4.0 keV, comparable to the temperature of the NGC~4839
  sub-group. No gas of  higher temperature, resulting from  a
  possible shock in the region of the Coma relic, is detected. 
Therefore, the connection between relics and cluster shocks seems to be
disproved by these data on the Coma cluster. The authors suggest
  that turbulence may be the major mechanism responsible for the
energy of particles. In particular, 
during its infall onto the Coma cluster, the sub-group around NGC~4839
encounters a region of relativistic particles and magnetic fields. The
interaction between the ionized moving plasma and the magnetic field
would imply energy transfer from the ICM to the relativistic particles.
Data on more objects will be crucial to
test the models and fully understand the relic phenomenon.

\begin{figure}
\centerline{
\psfig{figure=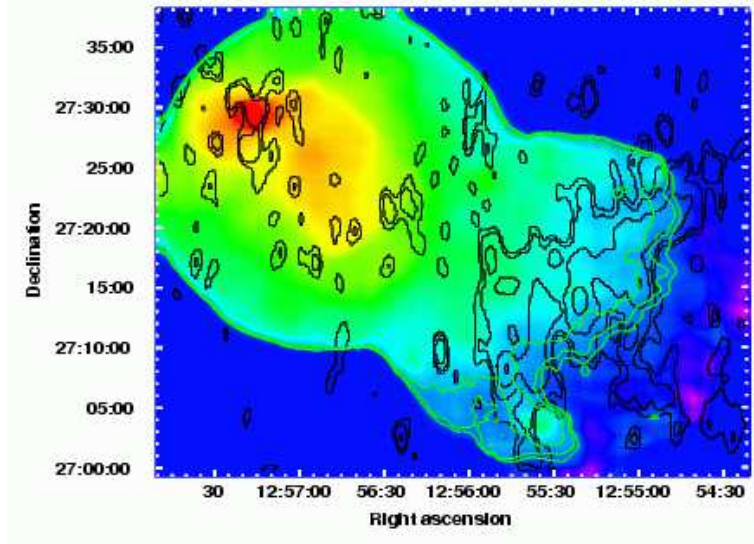,width=10cm}}
\caption
{
Color-scale X-ray emission of the NGC~4839 group obtained from the
XMM-Newton data (Feretti \& Neumann 2006).
Dark blue areas are regions with no X-ray data. The green contours represent
the X-ray brightness at 2, 3, and 5 $\sigma$ level over the background,
respectively. The radio emission from
the Coma cluster relic 1253+275 at 0.3 GHz is shown by the black contours.
}
\label{diff}
\end{figure}

\section {Conclusions}

Massive clusters of galaxies showing strong dynamical activity and
merger processes can host diffuse radio emission, which demonstrates
the existence of relativistic particles and magnetic fields in the
ICM.
From the comparison between radio and X-ray emission there is evidence
that recent merger phenomena would provide the energy for the
relativistic electron reacceleration, thus allowing the production of
a detectable diffuse radio emission.
A question which is still unanswered is
whether all merging clusters have cluster-wide radio halos. This
will be answered by systematic deep studies with future instruments.

Spectral index maps of the halos in A~665, A~2163 and A~2744 show that
 spectra are flatter in regions more directly influenced by the
 merger activity, confirming that electrons are likely gaining energy
 from the merger process.  Observational results favour a scenario
 where cluster turbulence might be the major responsible for the
 electron reacceleration in radio halos.

Theoretical models predicting a connection between cluster shocks and
radio relics need to be tested through X-ray observations of cluster
peripheral regions. In the Coma cluster, no clear shock is detected
at the location of the radio relic. More data on a significant
number of clusters, however, are needed to draw firm conclusions.

\section*{Acknowledgments}
I wish to thank the organizers for the very interesting meeting.
This work was partially supported by INAF under PRIN-2005.


\section*{References}
\begin{thebibliography}{99}

\bibitem{}
Bacchi M., Feretti L., Giovannini G., Govoni F., 
A\&A  400, 465 (2003)

\bibitem{}
Ballarati B., Feretti L., Ficarra A., Giovannini G., Nanni M., Olori C.,
Gavazzi G.,  A\&A 100, 323 (1981)

\bibitem{}
Blasi P., Jour. Kor. Astr. Soc. 37, 483 (2004)

\bibitem{}
Brunetti G., Setti G., Feretti L., Giovannini G.,
MNRAS 320, 365 (2001)

\bibitem{}
Brunetti G., Blasi P., Cassano R., Gabici S.,
MNRAS 350,  1174 (2004)

\bibitem{}
Buote D.A., ApJ 553, L15 (2001)

\bibitem{}
Cassano R., Brunetti G., MNRAS 357, 1313
(2005)

\bibitem{}
Cassano R., Brunetti G., Setti G., MNRAS 369, 1577 (2006)

\bibitem{}
En{\ss}lin T.A., Gopal-Krishna, A\&A 366, 26 (2001)

\bibitem{}
En{\ss}lin T.A., Br{\" u}ggen M.,  MNRAS 331, 1011 (2002)

\bibitem{}
Feretti L., Fusco-Femiano R., Giovannini G., Govoni F., 
A\&A 373, 106 (2001)

\bibitem{}
L. Feretti, in: {\it The Universe
at low radio frequencies}, IAU Symp. 199, ASP Conference Series,
Vol. 199,  p. 133 (2002)

\bibitem{}
Feretti L., Orr\'u E., Brunetti G., Giovannini G.,
Kassim N., Setti G., A\&A  423, 111  (2004)

\bibitem{}
Feretti L., AdSpR 36, 729 (2005)
%Cospar

\bibitem{authors}
Feretti L., Schuecker P., B\"ohringer H.,
Govoni F., Giovannini G.,  A\&A 444, 157 (2005)

\bibitem{}
Feretti L., Neumann D.M., A\&A 450, L21 (2006)

\bibitem{}
Fujita Y., Takizawa M., Sarazin C.L.,
ApJ 584, 190 (2003)

\bibitem{}
Gabici S., Blasi P.,  ApJ 583, 695 (2003)

\bibitem{}
Giovannini G., Feretti L.,
in: {\it Merging Processes of Galaxy
Clusters}, ASSL, Kluwer Ac. Publish., p. 197 (2002)


\bibitem{}
Giovannini G., Feretti L., Jour. Kor. Astr. Soc. 37, 323 (2004)



\bibitem{}
Govoni F., En{\ss}lin T.A., Feretti L., Giovannini G., 
A\&A 369, 441 (2001)


\bibitem{}
Govoni F., Markevitch M., Vikhlinin A., VanSpeybroeck L.,
Feretti L., Giovannini G., ApJ 605, 695
(2004)

\bibitem{}
Henry P.J., Finoguenov A., Briel U.G., ApJ 615, 181 (2004)

\bibitem{}
Hoeft M., Br\"uggen M., Yepes G., MNRAS 347, 389 (2004)

\bibitem{}
Keshet U., Waxman E., Loeb A.,  ApJ 617, 281
(2004)

\bibitem{}
Markevitch M., Vikhlinin A., ApJ 563, 95 (2001)
%shock a665 a2163


\bibitem{}
Markevitch M., Vikhlinin A., Forman W.R.,
in: {\it Matter and energy in clusters of galaxies},
ASP Conference Series  Vol. 301,  p. 37 (2003a)

\bibitem{}
Markevitch M., Mazzotta P., Vikhlinin A., ApJ 586, L19 (2003b)
%a754

\bibitem{}
Markevitch M., Govoni F., Brunetti G., Jerius D., ApJ  627, 733 (2005)

\bibitem{}
Orr\'u E.,
Feretti L., Govoni F., Murgia M., Giovannini G., Brunetti G., 
Setti G.,  AN 327, 565 (2006)

\bibitem{} 
Petrosian V.,  ApJ  557, 560 (2001)

\bibitem{}
Ricker P.M., Sarazin C.L., ApJ 561, 621 (2001) 

\bibitem{}
Ryu D., Kang H., Hallman E., Jones T.W.,  ApJ 593, 599 (2003)

\bibitem{}
Schuecker P.,  B\"ohringer H., in:  {\it Diffuse thermal
and relativistic plasma in galaxy clusters},  MPE Report 271, p. 43 (1999)

\bibitem{}
Schuecker P.,  B\"ohringer H., Reiprich T.H., Feretti L.,
 A\&A 378, 408 (2001)

\bibitem{} 
Willson M.A.G,  MNRAS 151, 1 (1970)


\end{thebibliography}
\end{document}